# First data and preliminary experimental results from a new Doppler Backscattering system on the MAST-U spherical tokamak

P. Shi,[a,b] R. Scannell,[a,1] J. Wen,[b] Z.B. Shi,[b] C. Michael,[c] T. Rhodes,[c] V.H. Hall-Chen,[d] Z.C. Yang,[b] M. Jiang,[b] and W.L. Zhong[b]

[a] *United Kingdom Atomic Energy Authority, Culham Science Centre,*
  *Abingdon, Oxon OX14 3DB, United Kingdom*

[b] *Southwestern Institute of Physics,*
  *PO Box 432, Chengdu 610041, People's Republic of China*

[c] *Department of Physics and Astronomy, University of California,*
  *Los Angeles, CA 90095, USA*

[d] *Institute of High Performance Computing, A*STAR,*
  *Singapore 138632, Singapore*
  *E-mail*: rory.scannell@ukaea.uk

ABSTRACT: A new Doppler backscattering (DBS) system, consisting of Q-band and V-band, has been installed and achieved its first data on the MAST-U spherical tokamak. The Q-band and V-band have separate microwave source systems, but share the same optical front-end components. The Q-band and V-band sources simultaneously generate eight (34, 36, 38, 40, 42, 44, 46 and 48 GHz) and seven (52.5, 55, 57.5, 60, 62.5, 65 and 67.5 GHz) fixed frequency probe beams, respectively. These frequencies provide a large range of radial positions from the low-field-side edge plasma to the core, and possibly to the high-field-side edge, depending on the plasma conditions. The quasi-optical system consists of a remotely-tunable polarizer, a focusing lens and a remotely-steerable mirror. By steering the mirror, the system provides remote control of the probed density fluctuation wavenumber, and allow the launch angle to match the magnetic field. The range of accessible turbulence wavenumbers ($k_\theta$) is reasonably large with normalized wavenumber $k_\theta \rho_s$ ranging from <0.5 to 9. The first data acquired by this DBS system is validated by comparing with the data from the other DBS system on MAST-U (introduced in Ref. [21]). An example of measuring the velocity profile spanning from the edge to the center in a high-density plasma is presented, indicating the robust capabilities of the integrated Q-band and V-band DBS systems.

# 1. Introduction

Anomalous transport induced by turbulence has emerged as the main contribution to cross-field particle and heat transport in tokamaks [1,2], directly influencing the performance and efficiency of fusion reactors. In the pursuit of controlled nuclear fusion, understanding and controlling turbulence is crucial for future fusion devices.

In tokamak experiments, Doppler backscattering (DBS) [3,4], also known as Doppler reflectometry, is a prevalent diagnostic tool for studying turbulence and transport. The DBS diagnostic provides a non-invasive means to measure density fluctuations and perpendicular flow velocities via the backscattering of a millimeter-wave beam launched obliquely toward a cutoff layer. The DBS technique operates on the principle of the Bragg condition, where the relationship between the probed density fluctuation, $k_{\tilde{n}}$, and the incident wave, $k_i$, at the scattering location is given by $k_{\tilde{n}} = -2k_i$. This condition constrains the backscattering process, allowing DBS to probe density fluctuations with a specific wavenumber, $k_{\tilde{n}}$, which depends on the wavenumber and launch angle of the incident beam. Notably, the DBS signal is conventionally thought to predominantly com from the cutoff layer, where $k_{\tilde{n}}$ is at its smallest value. Through the analysis of the Doppler frequency shift present in the backscattered turbulence signal, one can ascertain the turbulence flow velocity in lab frame. This can subsequently be converted into a localized $E \times B$ velocity, assuming that the $E \times B$ velocity is much larger than the turbulence velocity in plasma frame. Therefore, the DBS diagnostic facilitates localized measurements of both density fluctuations and $E \times B$ velocity.

In the past decades, a multitude of magnetic fusion devices including tokamaks and stellarators, such as MAST [5], DIII-D [6], ASDEX-U [7], KSTAR [8], EAST [9], HL-2A [10], Tore Supra [11], TJ-II [12] and LHD [13], have been equipped with DBS diagnostics. In these devices, the DBS diagnostics have empowered researchers to observe a diversity of turbulence-related phenomena, such as Trapped Electron Modes (TEM) [14], Electron Temperature Gradient (ETG) modes [15], Geodesic Acoustic Modes (GAM) [16], zonal flows [17], Ion Cyclotron Emission (ICE) [18], Alfvén modes [19]. The widespread adoption of the DBS technique in plasma devices globally, and its versatility in facilitating a diverse of measurements including – from low to high-k fluctuations, to GAMs and zonal flows, and energetic particle driven coherent modes – underscores its instrumental role in advancing our understanding of plasma instabilities. The ITER tokamak also plans to use DBS diagnostic [20].

Recently, a new DBS system has been installed and achieved its first data on the Mega-Ampere Spherical Tokamak Upgrade (MAST-U). It is worth noting that this DBS system is distinct from the other DBS system on MAST-U in Ref. [21], which is developed by the University of California (UCLA) team. This new DBS system is developed by the Southwestern Institute of Physics (SWIP). For clarity's sake, the DBS system from Ref. [21] will be referred to as the UCLA DBS, while the new DBS system presented in this paper will be referred to as the SWIP DBS. This SWIP DBS is comprised of a Q-band and a V-band system. While both SWIP DBS systems have separate microwave source systems, they utilize the same optical components. The Q-band system simultaneously launches eight fixed frequency (34, 36, 38, 40, 42, 44, 46 and 48 GHz) probe beams. The V-band source generates seven fixed frequency (52.5, 55, 57.5, 60, 62.5, 65 and 67.5 GHz) beams simultaneously. The launch angle in both poloidal and toroidal directions can be adjusted remotely by steering the mirror affixed to a rotating step motor. Section 2 will present the system design in detail. The first data and preliminary experimental results will be introduced in Section 3. Finally, a brief summary is included in Section 4.

# 2. System design

Benefiting from the broad range of launching frequencies (34~67.5GHz), the combined capabilities of the Q-band and V-band systems provide expansive radial coverage across diverse plasma scenarios, such as Ohmic low density, Ohmic high density, neutral beam injection (NBI) assisted low density H-mode, and NBI assisted high density H-mode plasmas. Figure 1 plots the cutoff profile for both O-mode ($f_{pe}$) and X-mode ($f_{rh}$) measurements, derived from four distinct actual plasmas on MAST-U. In O-mode operation, the Ohmic plasmas are transparent for most channels of V-band, the cutoff layer locations in H-mode plasmas are close to X-mode measurement. Thus, the SWIP DBS system consistently operates in X-mode. Interestingly, in X-mode operation, the V-band DBS is able to access to the very center of Ohmic plasmas, and can measure the top of pedestal in H-mode plasmas, even for very high density.

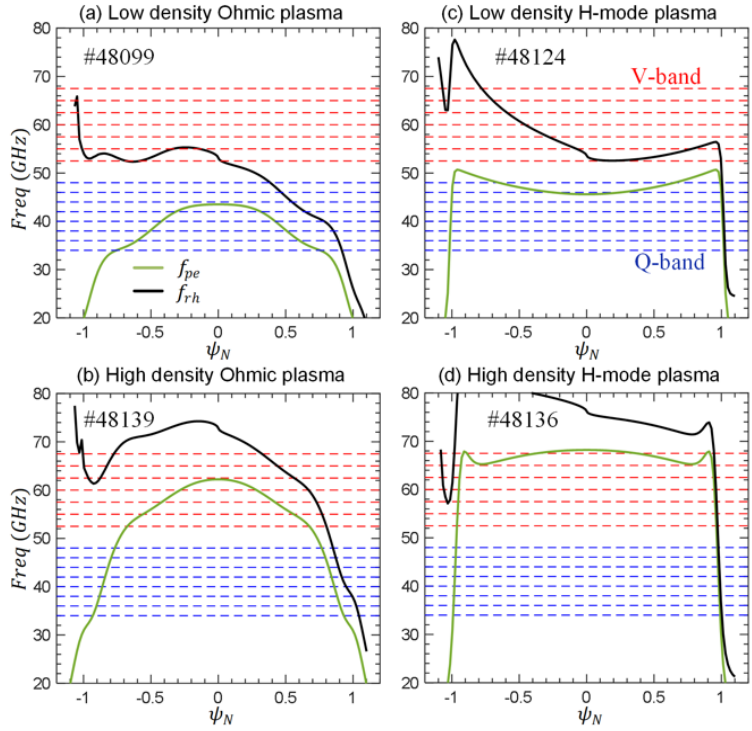

**Figure 1.** Four examples of radial access at midplane for the eight-channel Q-band and seven-channel V-band DBS systems on MAST-U. (a) Ohmic low density, (b) Ohmic high density, (c) Low density NBI H-mode, and (d) High density NBI H-mode.

In designing the microwave generating sources, SWIP incorporated existing Q-band and V-band DBS launch and receive systems. The Q-band utilizes a homodyne design, similar to the one implemented on HL-2A tokamak, as detailed in Ref. [10]. The V-band source is same to the system presented in Ref. [22], which is heterodyne and consisting of radio-frequency (RF) and intermediate-frequency (IF) systems. In addition, the multi-channel comb-frequency for both the Q-band and V-band sources are generated by the method of multiplexer-based frequency array source (MFAS) [23]. Figure 2 shows the layout of the DBS system, which is located at Sector 4 of MAST-U machine. Fig. 2 (a) and (b) depict the top view of the entire system and a cross-section view of the optical components, respectively. The optical system is composed of polarizer, focusing lens and a pair of mirrors. The polarizer is remotely controllable, for allowing the polarization angle of launch beam to match the pitch angle of magnetic field. The focusing lens is made of high-density polyethylene (HDPE), with a focal length of $35\ cm$. The mirror set includes a fixed mirror and a rotatable mirror. Beam steering, achieved by rotating the last mirror, determines the poloidal and toroidal launch angles of the DBS system. The range of launch angle for the poloidal and toroidal directions is $0\sim12°$. The angle of the last steering mirror is remotely set using two rotating step motors controlling the poloidal and toroidal mirror angles.

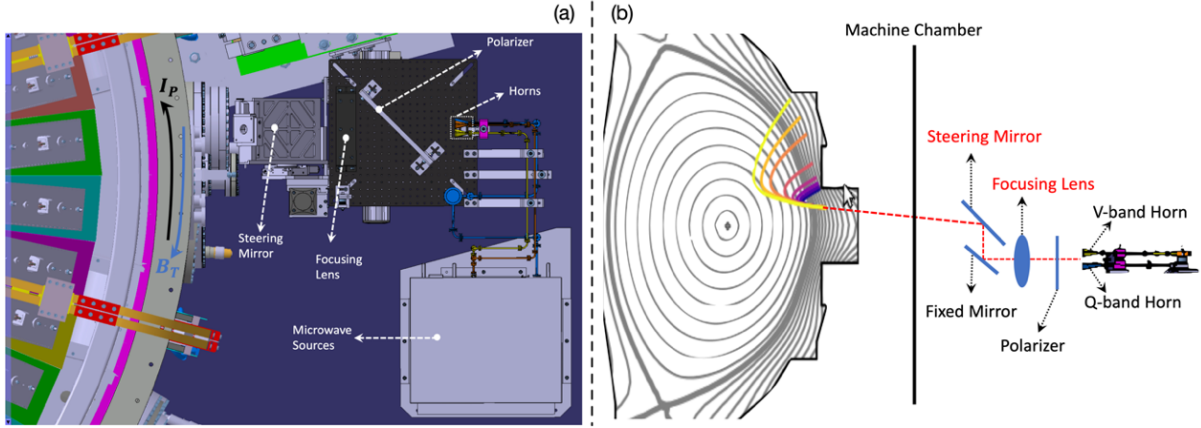

**Figure 2.** (a) Top view of the MAST-U and SWIP DBS systems, showing the relative locations of components of SWIP DBS system, and the directions of plasma current and toroidal field. (b) Schematic view of the cross section of plasma surfaces and optical components of the DBS system, illustrating the beam path from the horn to plasma.

Due to the space constraints, the antennas of Q-band and V-band cannot be aligned to the optical axis of the focusing lens simultaneously, as showed in Fig. 2(b). This arrangement results in an offset of the launch angles. The offset angles are $[\theta = 1.2°, \varphi = 0]$ for Q-band and $[\theta = -1.6°, \varphi = -2°]$ for V-band. Within this context, a positive poloidal angle indicates an upward direction in Fig. 2(b), while a positive toroidal angle represents a downward direction in Fig. 2(a). These offset angles are intentionally designed, to make the toroidal angle matching the poloidal angle being satisfied on Q-band and V-band simultaneously. The principle of how the mismatch angle affects the DBS signal is discussed in Ref. [24]. Based on the directions of plasma current and toroidal field on MAST-U, a positive toroidal launch angle matches a positive poloidal angle. Given a certain poloidal launch angle, a beam with a higher frequency necessitates a smaller toroidal launch angle, as indicated by Fig. 4 in Ref. [21]. Thus, the V-band DBS requires a smaller toroidal launch angle compared to the Q-band. If the Q-band launch angle is set as $[\theta = 6°, \varphi = 4°]$ by steering the mirror, then the V-band launch angle becomes $[\theta = 3.2°, \varphi = 2°]$. As such, both Q-band and V-band DBS can meet the matching condition.

With the launch angle range from 0 to $12°$, the DBS system is capable of targeting the turbulence with normalized wavenumbers $k_\theta \rho_s$ ranging from ≤0.5 to 9, based on 3D GENRAY [25] ray-tracing simulations. This wavenumber range is relevant to an extensive variety of instabilities (such as ITG, TEM, ETG, ICE, and kinetic ballooning modes (KBM)).

Figure 3 shows the evolution of beam waist radius along as the beam propagates. The starting beam waists for Q-band ($40 GHz$) and V-band ($60 GHz$) are $1 cm$ and $0.68 cm$ respectively. The black lines represent the HDPE focusing lens, and the dashed line denotes the plasma edge. The beam waist-radius around plasma edge for Q-band and V-band, are approximately ~$8 cm$ and ~$7.7 cm$ respectively. Obviously, the beam is not very focusing. A limitation of this design is its compromised spatial resolution. Nevertheless, a notable advantage is that the beam size, as well as the amplitude of DBS signal, are independent with the plasma position.

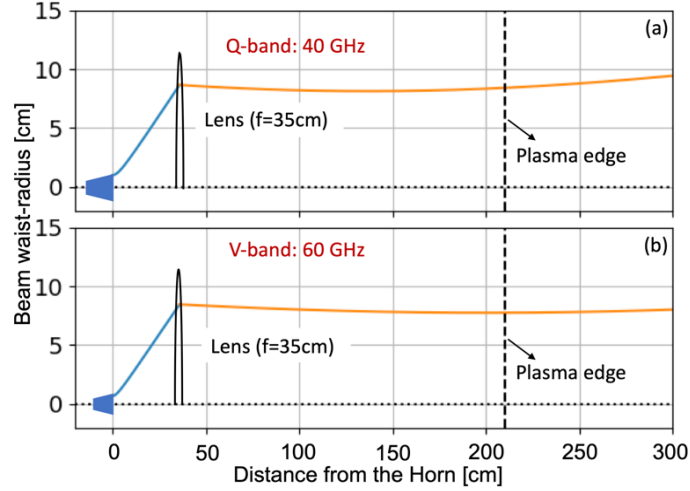

**Figure 3.** Beam width predictions from Gaussian beam calculations for (a) Q-band, $40 GHz$, and (b) V-band, $60 GHz$. The HDPE lens is represented by the black lines with focal length $f = 35\ cm$. The estimated plasma edge is indicated by the dashed lines.

## 3. Experimental results from SWIP DBS system

### 3.1 Data validation of the SWIP DBS system

As mentioned above, a DBS system developed by UCLA has been installed on MAST-U in 2021 [21]. Following an extensive scan in both toroidal and poloidal launch angles, we have gained confidence in the data from the UCLA DBS. Consequently, the UCLA DBS provides an excellent benchmark to validate the performance of the SWIP DBS. An example with the operation of the both DBS systems is shown in Figure 4. This is a NBI assisted L-mode plasma. And the two DBS systems launch at the same poloidal angle ($5°$), indicating that the Doppler shift should be the same if the probing frequencies are identical. From the Fig. 4 (c) and (d), it's evident that the density fluctuation measured by $40 GHz$ channel of SWIP DBS is almost the same to that of the UCLA DBS. Besides, the Doppler shift variation is likely due to the vibration of plasma density (Fig. 4 (b)).

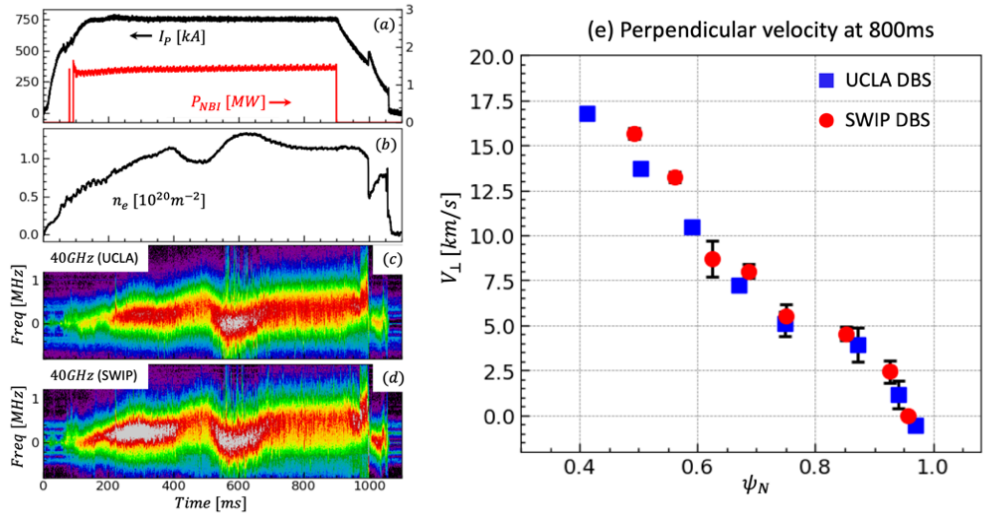

**Figure 4.** An example of off-axis NBI assisted L-mode plasma on MAST-U. (a) Total plasma current (black), and off-axis NBI power (red), (b) line-integral electron density by $CO_2$ interferometry, power spectrum of density fluctuation measured by $40 GHz$ probing beams on (c) UCLA DBS and (d) SWIP DBS, and (e) comparison of perpendicular velocities measured by UCLA DBS and SWIP DBS.

In addition, Fig. 4 (e) compares the perpendicular velocity profiles by the two DBS systems. The probing frequencies of UCLA DBS span from 32.5 to 50 $GHz$ in increments of 2.5 $GHz$, which is slightly different from the SWIP DBS. Nevertheless, the velocity profile measured by SWIP DBS is consistent with the result of UCLA DBS, further validating the data of SWIP DBS. It is noteworthy that a positive velocity corresponds to the ion diamagnetic direction, which is consistent with the direction of NBI driven toroidal velocity (co-current).

## 3.2 Measurement of Ohmic high-density plasma

As indicated in Fig. 1 (b), the V-band DBS can access the central region in high-density plasma. Figure 5 presents an example of Ohmic plasma on MAST-U, which disrupts around $480ms$ due to density limit. The maximum normalized Greenwald limit density is $\sim 1.4 n_G$. As showed in Fig. 5 (c) and (d), the DBS signals from Q-band ($40GHz$) and V-band ($60GHz$) experience a marked surge when the plasma density exceeds certain critical values. This is likely due to the outward displacement of the cutoff layers as the plasma density increases.

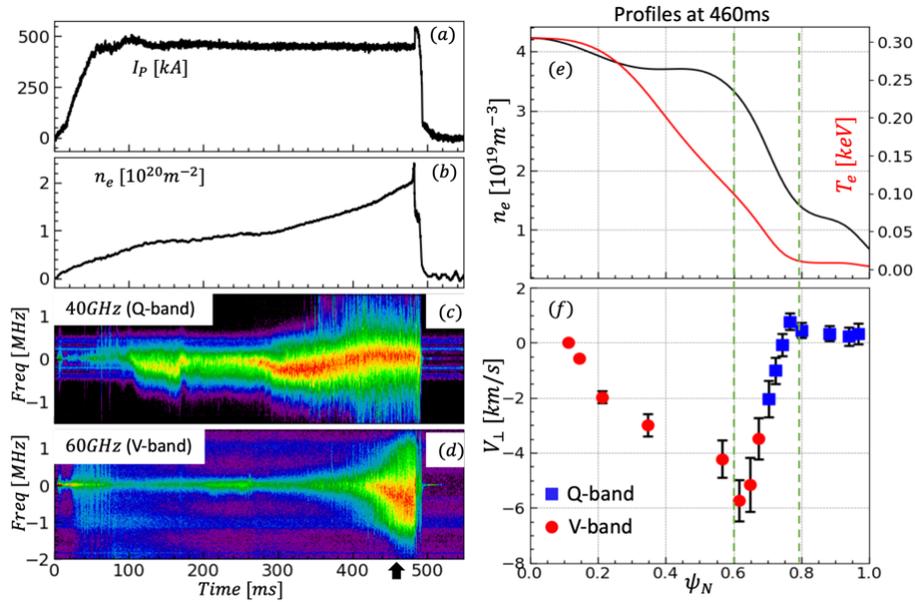

**Figure 5.** An example of high-density Ohmic plasma on MAST-U. (a) Total plasma current, (b) central line-integral electron density by interferometry, power spectrum of density fluctuation measured by probe beams of (c) Q-band $40GHz$ and (d) V-band $60GHz$, (e) electron density (black) and temperature (red) profiles by TS, and (f) perpendicular velocity profile measured by Q- and V-band DBS.

Fig. 5 (e) and (f) plot the profiles of electron density and temperature measured by Thomson scattering (TS), and perpendicular velocity obtained from the SWIP DBS system, for the time slice at $460ms$. The electron temperature profile indicates that the plasma is shrinking and a cool region is pushed into $\psi_N = 0.8$ preceding the density limit disruption. Meanwhile, the perpendicular velocities in the range of $\psi_N = 0.8 \sim 1$ are almost zero, suggesting an absence of flow shear in this region. The minimal flow shear at edge aligns with the extremely low temperature and small pressure gradient. In addition, by means of the V-band, the $E_r$ well is observed at an inner location of $\psi_N = 0.6$. The pronounced flow shear spanning $\psi_N = 0.6 \sim 0.8$ is consistent with the steep density profile in the corresponding region. It is noteworthy that measuring the central region in high-density plasmas is typically very difficult in conventional tokamaks. This finding underscores the immense application potential of the combined Q-band and V-band DBS on MAST-U.

## 4. Summary

A new DBS diagnostic developed by the Southwestern Institute of Physics (SWIP), has been successfully installed on MAST-U. Initial data acquired from this DBS system has been analyzed, yielding preliminary experimental results. The DBS system consists of Q-band and V-band, enabling observations of the center in L-mode plasmas, as well as the top of pedestal in H-mode plasmas. By comparing with the data from the previously-installed DBS system developed by UCLA, the data from the newly-installed SWIP DBS is validated. Furthermore, a case study involving central measurements in density limit disruption plasmas underscores the powerful capabilities of the DBS system on MAST-U.

## Acknowledgments

This work is supported by the National MCF energy R&D Program of China (Grant Nos. 2022YFE03060002 and 2022YFE03060003) and National Natural Science Foundation of China (Grant No. 12175055). This work has been also partly supported by the EPSRC Energy Programme [grant number EP/W006839/1].